\documentclass[12pt]{elsart}

\usepackage{natbib}
\usepackage{graphicx}

\usepackage{amssymb}

\begin{document}

\begin{frontmatter}



\title{Gamma-Rays as Probes for the Multi-Dimensionality  of Type Ia Supernovae}


\author{P. H\"oflich}

\address{Dept. of Astronomy, University of Texas, Austin, USA}

\begin{abstract}
 We present $\gamma $-ray spectra for a set of Type Ia supernovae models.
Our study is based on a detailed Monte Carlo transport scheme for
both spherical and full 3-D geometries. Classical and new challenges
of the $\gamma $ ray astronomy are addressed. 
 We find that $\gamma $-rays are very suitable to reveal the structure
of the envelope and, thus, they allow to probe properties of the nuclear 
burning front and the progenitor, namely its central density and global asphericities.
 The potential problems are discussed for the quantitative comparison between
theoretical and observed line fluxes during the first few months after the explosion.

\end{abstract}


\end{frontmatter}

\section{Introduction}
 $\gamma$-ray observations have long been recognized as a potential, valuable tool for
supernovae research \citep{clayton69,as88,chan91}.
 Only $\gamma$-rays provide a direct link to the Ni distribution which hardly depends on details
of the physics and on the numerical treatment.
Different scenarios can be distinguished by line fluxes and profiles, the structure of the
progenitors can be probed, and  the time of the explosion can be determined.
 $\gamma $-rays can provide a good determination of the $^{56}Ni$ production
for nearby SNe~Ia because, nowadays, accurate distances of nearby galaxies can 
be obtained by $\delta $Ceph.
  Moreover, all sky surveys by $\gamma $-rays may provide an unbiased rate of SNe~Ia.
As we will discuss below, 
advances in the fields of optical and IR observations and of the theory
have helped to redefine the goals of SN research by $\gamma $-rays.
New challenges emerged which emphasize their central role
as probes for the  3-D structure of SNe~Ia.

 The results presented are based on our gamma-ray codes
\footnote{available on request} for spherical
\citep{hmk92} and arbitrary 3-D geometries \citep{hl01}. We assume
homologous expansion of  density and chemical structures calculated by 
spherical and 3-D hydro simulations.
 All nuclear decay lines of $^{56}Ni$ and $^{56} Co$ are included.
 Pair production and bound-free opacities are taken into account.

\section{Classical and New Challenges}
\noindent{\bf Classical questions and problems:} In parts,
this section is based on our previous analyses \citep{hmk92,hwk98}.
 For further discussions, we also want to refer to \citet{burrows91},\citet{kumagai97} and
\citet{pinto01}.
\begin{figure}
\includegraphics[width=9.2cm,angle=270]{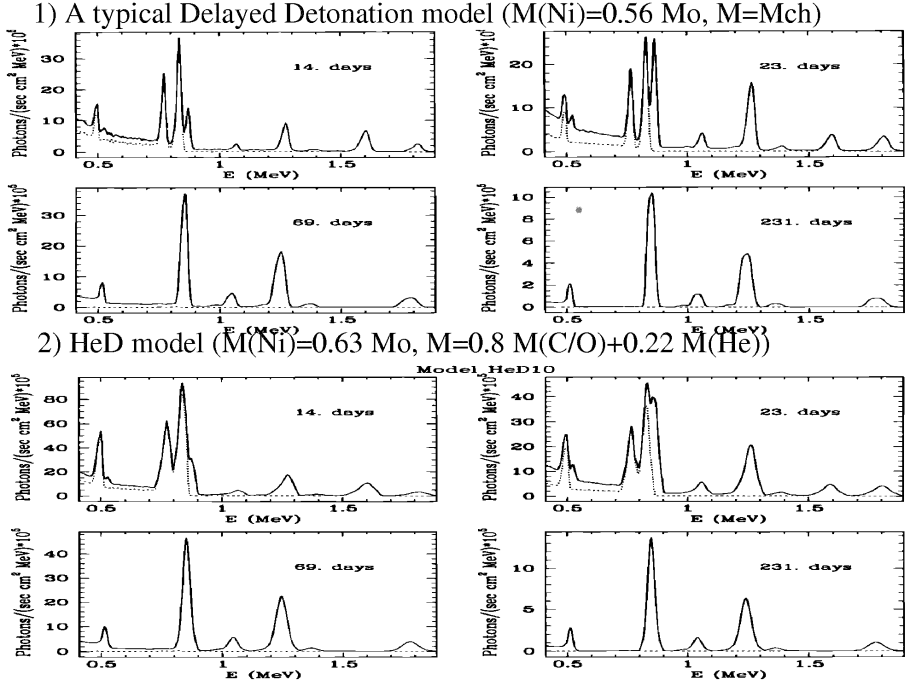}
\caption{
 Comparison of a typical delayed detonation model (DD201) with
a helium triggered Sub-Chandrasekhar model (HeD10) (from \citet{hwk98}).
The dashed lines correspond to the contribution of $^{56}Ni$.
}
\label{f1}
\end{figure}
\begin{figure}
\includegraphics[width=5.2cm,angle=270]{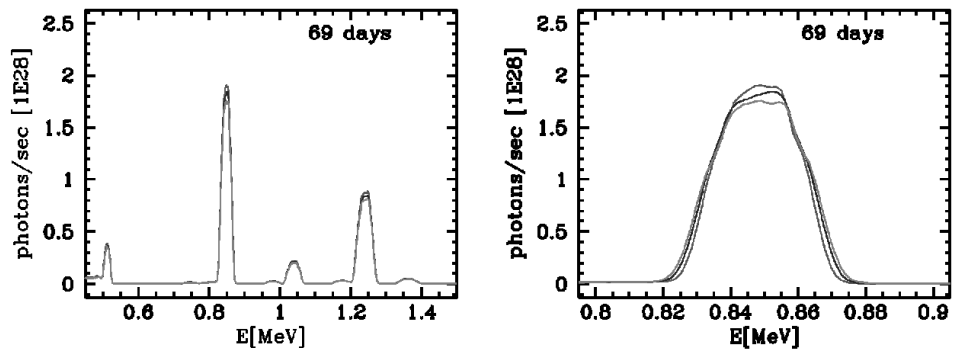}
\caption{ Influence of the central density on $\gamma$-ray spectra at the example of
a delayed-detonation model (Model $5p0z22.20$ of H\"oflich et al. 2001) 
for $\rho_c = 2, 4 \& 6 \times 10^9 g~cm^{-3}$ (top to bottom lines).
}
\label{f2}
\end{figure}
 Overall, $\gamma$ spectra and their evolution is characterized by a turnover from a phase
dominated by $^{56}Ni$ to $^{56}Co$ lines (Fig. \ref{f1}). The time of the explosion can be
determined by the  ratio between the $^{56}Ni$(0.81MeV) and the $^{56}Co$(0.84 MeV) lines because it
varies strongly with time of the explosion but it hardly depends on the model.
 With time, the envelope
becomes increasingly transparent. The spectral evolution depends sensitively on
the density and chemical structure of the envelope and, thus, provides a valuable tool
for the discrimination of explosion models. For example,
 sub-Chandrasekhar mass models show as  a distinguishing  feature 
 an outer layer of $^{56}Ni$ which reveals itself by high $\gamma$-ray fluxes and broads
 already a few days after the explosion 
(Fig. \ref{f1}). At late phases ($\geq $100 days), the absolute line fluxes
 and  profiles  provide a direct measure of the total $^{56}Ni$ mass and its distribution.
For $M_{Ch}$ models, the high densities close to the center ($\geq  10^9 g~ cm^{-3}$) result in the production of
neutron-rich iron-group isotopes rather than
 $^{56}Ni$.
The size of this central region increases with $\rho_c$. 
A knowledge of this property is critical to detect  because systematic variations in $\rho_c$
 can produce an offset (up to $0.2 ^m$, \citet{dominguez01})
in the  brightness decline relation \citep{hamuy95},
a cornerstone of modern cosmology with SNe~Ia (e.g. \citet{schmidt98,perlmutter99}).
 Optical and near IR-spectra do not allow to  distinguish isotopes.
 However, the lack of central $^{56}Ni$ reveals itself in by flat-topped line profiles (Fig.
\ref{f2}).
 To detect the variation, we need resolutions between 20 to 30 which are well within reach for the
 upcoming INTEGRAL mission.

 We want to mention one problem related to the analysis of observed line fluxes during
the first months after the explosion.
An advantage of $\gamma $-ray compared to optical analyses is that
the results are insensitive to details
of the physics or numerical treatment. E.g. the specific
opacities do not depend on temperature or density. This advantage is somewhat lost along the way
when comparing the observations with  theoretical predictions.
Problems are caused by the time- and model-dependent line shifts and widths ($\approx 10,000 km/s$),
the intrinsic response function of the instrument which is  non-Gaussian,
and they are connected to the actual definition used to determine integrated
 line fluxes from synthetic spectra.
The actual value  may differ by up to a factor of 2 even if based on a given synthetic spectrum
(e.g. \citet{mueller91}). Preferable, the same machinery should be employed
 for both the synthetic spectra and the observations to derive accurate values or good
 upper limits for the $\gamma$ emission.

\begin{figure}
\includegraphics[width=5.2cm,angle=270]{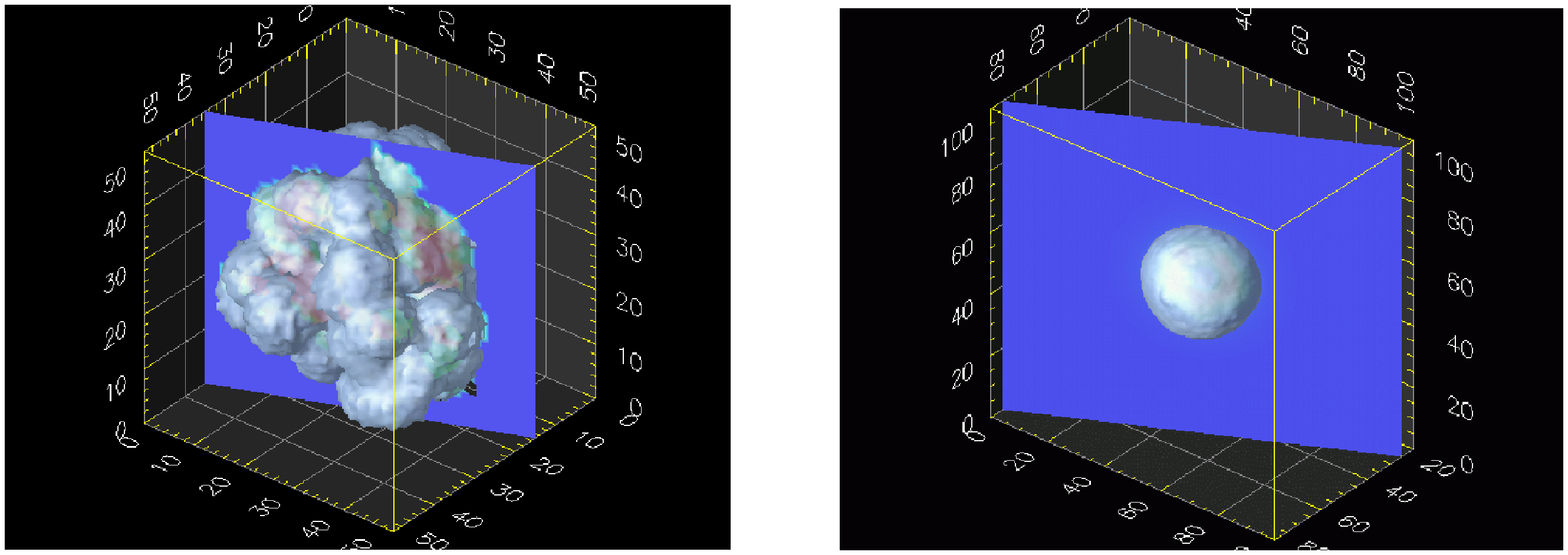}
\includegraphics[width=5.3cm,angle=270]{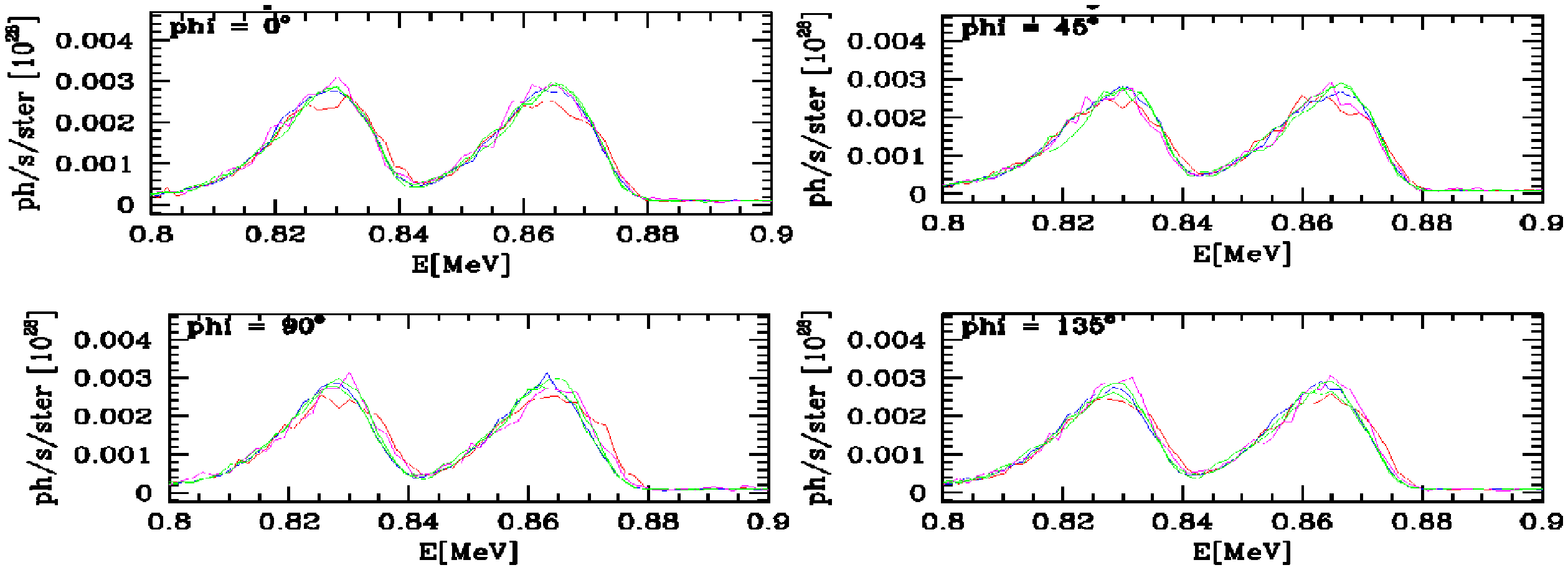}
\caption{ 
Energy deposition by $\gamma $-rays at day 1 (left) and 23 (right)
based on our full 3-D MC gamma ray transport \citep{hl01} (upper panels). The diameter
of the WD is normalized to 100. In the lower panels, the
 $\gamma $ spectra at day 23 are given as seen from various $\Theta $ of
 90, 60, 30 and 0$^o$ (red, blue, green, pink) and $\phi$.
In the presence of inhomogeneities, typical fluctuations of about 10 \%
can be seen. Their frequency provides a measure of the scale.
The explosion model is based on the delayed detonation model 5p0z22.20 (\ref{f2})
with $\rho_c = 2 \times 10^9 g~cm^{-3}$ 
assuming chemical inhomogeneities according to \citet{khokhlov01}.
}
\label{f3}
\end{figure}

\noindent{\bf New challenges:}
During the last few years,  observational and theoretical methods in the field of SN-research advanced
significantly  well beyond the point imaginable as little as 10 years ago. Back then,
 SNe~Ia were discovered a few days before maximum or later and, in general, light curves were rather
uncertain and the spectral coverage was poor. Even for nearby SNe~Ia, distances to the host galaxy
were uncertain by 20 to 30 \% . Now, accurate $\delta $-Ceph.
based distances of nearby galaxies are available (e.g. \citet{saha97}). Combined with
the establishment of the optical brightness decline relation relation \citep{hamuy95},
SNe~Ia provide a unique tool for cosmology. Robotic, systematic
 SN-search programs (e.g. LOSS, c.f. \citet{treffers97})
  continuously increase the sample both at low, intermediate and 
high red-shifts. Often, SNe~Ia are discovered 4 to 5 magnitudes before maximum light
 \citep{riess99,aldering00},
 almost eliminating the selection effects by galactic extinction and providing
a tight handle on the rise times.
 Detailed observations of optical and infrared spectra and light curves allowed sophisticated analyses
and test of scenarios, ruling all but out the once popular helium triggered
detonations, and strongly favoring deflagration or delayed detonation 
$M_{Ch}$ models or, in some instances, the merger scenario (e.g. \citet{hk96}).
 Observation of polarization in SNe~Ia has shown that, in general, these objects are fairly
spherical \citep{wang01} with the noticeable exception of very subluminous SNe~Ia \citep{howell01}.
 For the first time, a direct connection
with the progenitors seems to be within reach. In particular, there is mounting evidence for a
connection between the properties of the progenitor, and the physics of the explosion
\citep{hwt98,iwamoto99,dominguez01,khokhlov01}.
 The recent progress may redefine the role of modern $\gamma $-ray astronomy for the field of
supernovae, and some of classical goals may have been rendered less compelling.
 $\gamma $-rays are particular valuable to measure the 3-D structure of SNe~Ia.
A comprehensive list of new goals is beyond the format of this paper.
 We want to address two of the areas, namely, $\gamma $-rays as tools to reveal
properties of nuclear burning front and large scale asymmetries.

\noindent {\sl Nuclear burning fronts:}
 Within $M_{Ch}$ models, optical and IR LCs and spectra can be reproduced by
models in which a (slow) deflagration front turns into a detonation (e.g. \citet{khokhlov91})
or, alternatively, a deflagration front is rapidly accelerating as in W7  \citep{nomoto84}.
However, successful models require parameterized descriptions for the propagation of the burning front.
For a discussion, see \citet{dominguez01} and references therein.
 The propagation of a detonation front is well understood but the description of
the deflagration and the deflagration to detonation transition (DDT) pose 
problems.
On a microscopic scale, a deflagration propagates due to heat conduction by 
electrons. Though the laminar flame speed in SNe~Ia is well known, the front has
been found to be Rayleigh-Taylor (RT) unstable, increasing the effective speed 
of  burning \citep{nomoto76}. Recently, significant
progress has been made toward a better understanding of the physics of flames.
Starting from static WDs, hydrodynamic calculations of the deflagration fronts 
have been performed in 2-D \citep{lisewski00}, and full 3-D 
\citep{khokhlov01}.  These calculations  demonstrated the complicated morphology
of  the front.  \citet{khokhlov01} finds that, while the expansion of the
envelope becomes almost spherical, the inhomogeneous chemical structure will 
fill about 50 to 70\% of the star (in mass).
 The resulting  chemical inhomogeneities and their scale depends sensitively on 
 the structure of  the initial WD, i.e.  progenitor  and the pre-conditioning of the runaway.
 If a DDT occurs at densities
needed to reproduce normal-bright SNe~Ia, most of the unburned fuel  
 will be consumed during the detonation phase and, by enlarge, the
chemical inhomogeneities will be eliminated. However, they will survive in pure deflagration models.

 Currently, state of the art 3-D calculations are restricted to the regime of 
linear instabilities, i.e. the flamelet regime. The consistent treatment
is  restricted to the early part of the deflagration phase.
In these calculations, The initial acceleration of the
  the deflagration front is  followed by a
declining rate of burning. This  leave a significant fraction
of the WD unburned ($\approx 0.5 M_\odot$).
  The resulting structures  cannot account for the
observations of typical SNe~Ia \citep{khokhlov01}.
 This  problem is  well known  from spherical deflagration models with a low deflagration speed
(e.g. DF1, \citet{khokhlov91}), and it triggered the suggestion
of a DDT, or, alternatively, showed
 the need for continuously accelerating deflagration front \citep{nomoto84}.
Despite the limitations of current models, the chemical clumps will not mix in the subsequent phase.
 To test for possible effects of chemical inhomogeneities  produced during the deflagration,
we have remapped a 3-D  chemical structure on a spherical explosion model (Fig. \ref{f3}).
The resulting  line profiles show fluctuations of about 10 \%.
 Their frequency provides a measure of the scale of the instabilities which depends on the C/O ratio of the
progenitor. Weak or absent fluctuations in the $\gamma $-ray spectra would
be a strong indicator for a DDT.

\begin{figure}
\includegraphics[width=5.2cm,angle=270]{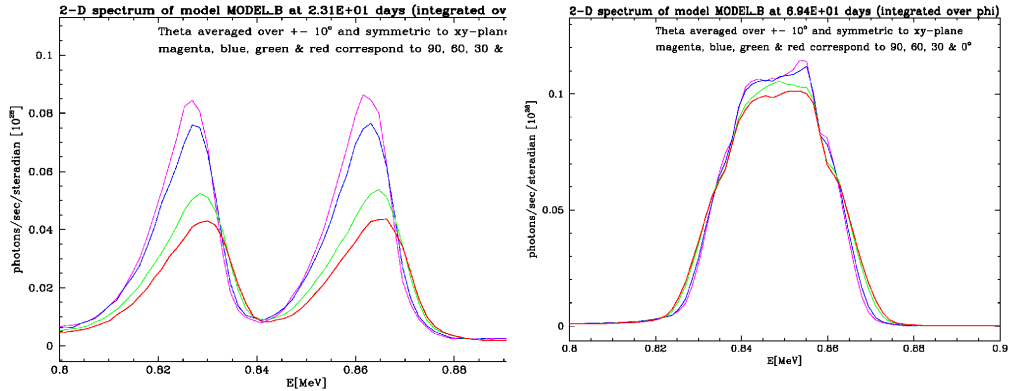}
\caption{ 
 Directional dependence of the spectrum of an ellipsoidal delayed detonation model (see Fig. \ref{f3})
at day 23 and 69 after the explosion. We assumed an axis ratio of 0.8 consistent with the
optical polarization data for SN1999by.
}
\label{f6}
\end{figure}

\noindent {\sl Large scale asymmetries:}
 One of the open questions is the nature of subluminous SNe~Ia such as SN~1991bg
\citep{filippenko92,leibundgut93}.
 Among them, SN~1999by is one of the best observed SNe~Ia. In addition to the studies 
 of optical light curves, detailed polarization spectra of the subluminous
 SN~1999by have been obtained and analyzed \citep{howell01}. Whereas `normal' SNe~Ia tend to show
little or no polarization \citep{wang97}, this supernova was
significantly polarized, up to 0.7\%, indicating an overall asphericity of the photosphere of
$\approx 20\% $.  This result suggests that there may be a connection between
the observed asphericity and the subluminosity in SNe~Ia. Among others, a possible explanations are
the explosion of a rapidly rotating WD and its effect on the propagation of
nuclear flames during the explosive phase of burning, or extensive burning of carbon 
just prior to the runaway \citep{hgfs01}. Polarization in the optical measures the
asymmetry of the photosphere but, in general,
 provides little information whether it is caused by a global asymmetry in the density structure or
in the excitation mechanism, i.e. the $^{56}Ni$ distribution.
 To test the sensitivity of $\gamma $-rays,
we have calculated the transport for the subluminous model of ellipsoidal shape
(Figs. \ref{f6}). The flux and line profiles vary as a function of
inclination (e.g. by $\approx 50 \%$ at day 23). In combination
with optical and IR observations, $\gamma $-rays can be a key to answer the question
on the nature of the asymmetry.

\end{document}